\documentstyle[aps]{revtex}
\begin{document}
\widetext
\preprint{preprint mpi-pks 9805014}
\title{Excitation Spectrum of the Holstein Model}
\author{J. M. Robin}
\address{Max-Planck-Institut f\"ur Physik komplexer Systeme,
N\"othnitzer Stra{\ss}e 38,
01187 Dresden, Germany}
\date{\today, e-mail: robin@mpipks-dresden.mpg.de}
\maketitle
\draft
\begin{abstract}
In this paper the polaron problem for the Holstein model is studied
in the weak coupling
limit. We use second order perturbation theory to construct renormalized
electron and phonons. Eigenstates of the Hamiltonian are labelled 
and the excitation spectrum is constructed. 
\pacs{ pacs numbers:  71.38.+i, 63.20.Kr}

\end{abstract}



\section{Introduction}
The Holstein model which describes electrons coupled to some local 
molecular deformations is a simple model to study 
polaron properties\cite{Holstein,Mott}. 
These recent years, accurate numerical 
investigations of this model on finite size systems have 
given new insights\cite{Mott,Thibblin,Stephan,Fehske,Kornio,White}
closely linked to some current problems 
in condensed matter physics\cite{Mott}.
A polaron corresponds to an electron renormalized by the interactions 
with the lattice deformations or phonons\cite{Josef}. 
From a historical point of view\cite{Mahan,LLP,Feynman} the approach to the 
polaron problem consists in computing the spectrum of the whole system 
containing a single electron coupled to some optical phonons. Writing $k$
the momentum of the electron and $Q$ the momentum of the phonons, the 
problem is to compute the lowest energy of the system for a  total momentum 
$K=k+Q$, since the total momentum operator commute with the Hamiltonian. 
This mimimum eigenenergy of the Hamiltonian for a given total momentum $K$
gives the polaron relation dispersion, as far as we consider only the low 
lying excitations of the system --that is as far as no phonon excitations 
are involved.

In this paper, we consider a different approach to the problem and try
to decompose the excitations of the system in term of polaron excitations
and renormalized phonons excitations. Due to the difficulty of the problem,
we address only the weak coupling limit of the problem and we use mainly 
second order perturbation theory in the electron phonon coupling. 
Within this framework, one can  block diagonalize the Hamiltonian.
Then it is possible to label the eigenstates of the Hamiltonian in term
of renormalized electron and phonons. 
In other words, an eigenstate of the Hamiltonian contains one renormalized
electron (a polaron) of momentum $k$ and some renormalized phonons of
momentum $Q$; the total momentum of the state being $K=k+Q$.

The Hamiltonian for the Holstein model, in Wannier representation, is given by
\begin{equation}
H \; = \; - t \sum_{j,\delta} c_{j+\delta}^{\dagger} c_{j} \; + \; 
\omega_{0} \sum_{j} b_{j}^{\dagger} b_{j} \; - \; g \omega_{0} \sum_{j}
c_{j}^{\dagger} c_{j} ( b_{j} + b_{j}^{\dagger} ) ,
\end{equation}
where $c_{j}^{\dagger}$ and $c_{j}$ are electron creation and annihilation 
operators, $b_{j}^{\dagger}$ and $b_{j}$ are phonon creation and annihilation 
operators. $t$ is the hopping integral, $\omega_{0}$ is the optical phonon
frequency and $g$ is the dimensionless coupling constant. 
In the strong coupling limit, $g \gg 1$, the Hamiltonian is approximatively
diagonalized by the Lang-Firsov transformation\cite{Lang-Firsov,Mahan}.
The transformed Hamiltonian is given by $\tilde{H} = e^{-S} H e^{S}$, with
$S = \sum_{j} g c_{j}^{\dagger} c_{j} ( b_{j}^{\dagger} - b_{j})$. It is 
only when the hopping term is zero that the transformed Hamiltonian is 
diagonal. In this case one obtains some new fermionic operators, which 
describe small polarons, or localised polarons. The small polaron 
creation operator is
given by $p_{j}^{\dagger} = e^{S} c_{j}^{\dagger} e^{-S}$, or 
$p_{j}^{\dagger} = U_{j} c_{j}^{\dagger}$, with 
$U_{j} = \exp g (b_{j}^{\dagger} - b_{j})$ while the renormalized phonons
operators are given by ${\cal B}_{j} = b_{j} - g$.
The action of this small polaron creation operator on the vacuum 
is thus to create an electron 
and a coherent state for the phonons where the annihilation operator takes
the expectation value $g$. 
For non zero values of the hopping integral, these small polarons become
some quasiparticles more or less well defined depending on the value
of the coupling and on the phonon frequency\cite{Mott,jm97}. 
We notice that for $g=0$, the small polaron correspond to the bare
electron, since there is no coupling and thus no renormalization.
We shall use a similar unitary transformation formalism but in we weak
coupling limit, $g \rightarrow 0$, and work in momentum space
where the Hamiltonian reads
\begin{equation}
H \; = \; \sum_{k} \varepsilon_{k} c_{k}^{\dagger} c_{k} \; + \; 
\omega_{0} \sum_{q} b_{q}^{\dagger} b_{q} \; 
- \; \frac{g \omega_{0}}{\sqrt{M}} \sum_{k,q} c_{k+q}^{\dagger} c_{k} ( 
b_{q} + b_{-q}^{\dagger} )
\end{equation}
where $M$ is the number of sites of the lattice, and indices of various
operators correspond to Bloch momenta. 
In the next section we consider our approach for a two site system, then 
we will generalize to a general system.

\section{The two site system}
We start our perturbation theory for the two site system. In this case, the
momentum takes two values $k=0,\pi$ and the bare electron relation dispersion
is given by $\varepsilon_{0}=-t$ and $\varepsilon_{\pi}=+t$. The $q=0$ mode
of the phonons is coupled to the total number of electron which is one in 
this problem. We drop out this mode after performing the shift corresponding
to the displaced oscillator\cite{Thibblin}. 
The reduced Hamiltonian is then given by
\begin{equation}
H_{R} \; = \; - \frac{g^{2} \omega_{0}}{2} ( c_{0}^{\dagger} c_{0} + 
c_{\pi}^{\dagger} c_{\pi} ) \; - t \; ( c_{0}^{\dagger} c_{0} - 
c_{\pi}^{\dagger} c_{\pi} ) \; + \; \omega_{0} b^{\dagger} b \; - \; 
\frac{g \omega_{0}}{\sqrt{2}} ( c_{0}^{\dagger} c_{\pi} 
+ c_{\pi}^{\dagger} c_{0} ) ( b + b^{\dagger} )
\end{equation}
We have dropped the index $\pi$ for the boson operators.
The basis states are given by 
$| k ; n \rangle = c_{k}^{\dagger} (b^{\dagger})^{n}/ \sqrt{n!} | 0 \rangle$
and $| 0 \rangle$ is the vacuum.
For $g=0$ the corresponding eigenenergies are given by 
$\varepsilon_{n}(k) = n \omega_{0} + \varepsilon_{k} - g^{2}\omega_{0}/2$.
Using second order perturbation theory, we obtain the eigenstates and 
eigenenergies,
\begin{equation}
E_{n}(k) \; = \; \varepsilon_{n}(k) \; + \; \frac{g^{2} \omega_{0}}{2} \left[
n \frac{\omega_{0}}{2 \varepsilon_{k} + \omega_{0}} + (n+1) \frac{\omega_{0}}{
2 \varepsilon_{k} - \omega_{0}} \right]
\end{equation}
\begin{equation}
| k ; n ) \; = \; \left\{ 1 + \frac{g\omega_{0}}{\sqrt{2}} \left[ 
\frac{b}{2t - \omega_{0}} + \frac{b^{\dagger}}{2t + \omega_{0}} \right]
c_{\pi}^{\dagger} c_{0} \; - \; \frac{g\omega_{0}}{\sqrt{2}} \left[
\frac{b}{2t + \omega_{0}} + \frac{b^{\dagger}}{2t - \omega_{0}} \right]
c_{0}^{\dagger} c_{\pi} \right\} \; | k ; n \rangle
\end{equation}
We show now that the eigenstate $| k ; n )$ describes a state with one polaron
of momentum $k$ and $n$ renormalized phonons.
The states $| k , n )$ are eigenstates of the Hamiltonian $H_{R}$, thus
using $\tilde{H}_{R} = e^{-S} H_{R} e^{S}$, we obtain 
$| k ; n ) = e^{S} | k ; n \rangle$. Up to ${\cal O}(g^{2})$, the 
$S$-matrix of the unitary transformation is given by 
\begin{equation}
S \; = \; \frac{g\omega_{0}}{\sqrt{2}} \left[ 
\frac{b}{2t - \omega_{0}} + \frac{b^{\dagger}}{2t + \omega_{0}} \right]
c_{\pi}^{\dagger} c_{0} \; - \; \frac{g\omega_{0}}{\sqrt{2}} \left[
\frac{b}{2t + \omega_{0}} + \frac{b^{\dagger}}{2t - \omega_{0}} \right]
c_{0}^{\dagger} c_{\pi}
\end{equation}
We note that if we set $t=0$ in this expression, we obtain the right 
transformation which diagonalize the reduced Hamiltonian $H_{R}$. 
We can now construct the corresponding polaron operator $p_{k}$ through 
the transformation $p_{k} = e^{S} c_{k} e^{-S}$.
We obtain explicitly
\begin{equation}
p_{0} \; = \; c_{0} + \frac{g\omega_{0}}{\sqrt{2}} \left[ 
\frac{b}{2t+\omega_{0}} + \frac{b^{\dagger}}{2t-\omega_{0}} \right]
c _{\pi}
\end{equation}
\begin{equation}
p_{\pi} \; = \; c_{\pi} - \frac{g \omega_{0}}{\sqrt{2}} \left[ 
\frac{b}{2t-\omega_{0}} + \frac{b^{\dagger}}{2t+\omega_{0}} \right]
c_{0}
\end{equation}
These operators are fermionic (they satisfy the anticommutation relations),
with the property  
$p_{k}^{\dagger}p_{k} | k' ; n ) = \delta_{k,k'} | k ; n )$. This means
that the eigenstate $| k ; n )$ contains one polaron of momentum $k$. As
concern the phonons, we can proceed the same way, introducing a new operator
${\cal B} = e^{S} b e^{-S}$ which describes a renormalized boson or phonon. 
Explicitly, one obtains
\begin{equation}
{\cal B}^{\dagger} {\cal B} \; = \; b^{\dagger} b \; + \; 
\frac{g \omega_{0}}{\sqrt{2}} \left[ \frac{b^{\dagger}}{2t-\omega_{0}} -
\frac{b}{2t-\omega_{0}} \right] c_{0}^{\dagger} c_{\pi}  \; + \;
\frac{g \omega_{0}}{\sqrt{2}} \left[ \frac{b}{2t-\omega_{0}} - 
\frac{b^{\dagger}}{2t+\omega_{0}} \right] c_{\pi}^{\dagger} c_{0}
\end{equation}
Again, this operator is diagonal with 
${\cal B}^{\dagger} {\cal B} | k ; n ) = n | k ; n )$. 
We therefore conclude that the state $| k ; n )$ contains one polaron of 
momentum $k$ and $n$ renormalized phonons. The total momentum of the state is 
$K=k$ if $n$ is even and $K=k+\pi$ if $n$ is odd. Then we write the 
eigenenergies as $E_{n}(k) = \varepsilon_{n}^{*} \pm t_{n}^{*}$. One 
obtains
\begin{equation}
\varepsilon_{n}^{*} \; = \; n \omega_{0} - \frac{g^{2} \omega_{0}}{2}
\frac{4t^{2} - 2 \omega_{0}^{2}}{4t^{2} - \omega_{0}^{2}}
\end{equation}
\begin{equation}
t_{n}^{*} \; = \; t \left\{ 1 + g^{2} \frac{\omega_{0}^{2}}{4t^{2} - 
\omega_{0}^{2}} (2n+1) \right\}
\end{equation}
For $n$ phonons or renormalized phonons in the system $t_{n}^{*}$ 
corresponds to the fermionic (polaronic) excitation energy.
In the polaron problem, we restrict ourselves to the case $n=0$, that is
we create a polaron from the vacuum,
$| k ; 0 ) = p_{k}^{\dagger} | 0 \rangle$.
We are left with $\varepsilon^{*} = \varepsilon_{0}^{*}$ which is the polaron
binding energy and $t^{*} = t_{0}^{*}$ which represents the hopping
integral of the polaron.
If we take $g=0.4$ and $\omega_{0} = 10/11t$ we obtains
$\varepsilon^{*} = -0.0538t$ and $t^{*} = 1.042t$ while exact
diagonalization\cite{Thibblin} gives $\varepsilon^{*} = -0.0576t$ and
$t^{*} = 1.038t$. We notice the enhancement of the bandwidth for this
two site system.

In term of the renormalized operators, the Hamiltonian can now be written
in a block diagonal form,
\[
H \; = \; \sum_{k} E_{0}(k) p_{k}^{\dagger} p_{k} \; + \;
\frac{1}{2!} \sum_{k} [ E_{1}(k) - E_{0}(k) ] {\cal B}^{\dagger} {\cal B}
p_{k}^{\dagger} p_{k} 
\]
\begin{equation}
\mbox{\hspace{2cm}} \; + \;
\frac{1}{3!} \sum_{k} [ E_{2}(k) - 2 E_{1}(k) + E_{0}(k) ]
{\cal B}^{\dagger} {\cal B}^{\dagger} {\cal B} {\cal B} p_{k}^{\dagger} p_{k}
\; + \; \ldots
\end{equation}
The original electron phonon interaction which contained off diagonal 
transitions between subspaces with the same number of particle (electron or
phonon) has been eliminated:
there is no more off diagonal 
interactions between the polaron and the renormalized phonons.

\section{General Case}
Next, we consider finite size systems for which numerical results 
have been obtained\cite{Stephan,Fehske}. 
Let $| \vec{n} \rangle$ be the state with the configuration of phonons 
$\{ \vec{n} \}$ such that 
$b_{q}^{\dagger} b_{q} | \vec{n} \rangle = n_{q} | \vec{n} \rangle$.
The basis vector are then 
$| k; \vec{n} \rangle = c_{k}^{\dagger} | \vec{n} \rangle$. The 
eigenstates are given by $| k; \vec{n} ) = e^{S} | k; \vec{n} \rangle$.
Now we use the fact that the state $| \vec{n} \rangle$ contains no 
electron, so that $e^{-S} | \vec{n} \rangle = | \vec{n} \rangle$. 
Using the definition of the polaron operator 
$p_{k} = e^{S} c_{k} e^{-S}$,  we obtain the desired relation,
$p_{k}^{\dagger} | \vec{n} \rangle = | k ; \vec{n} )$. 
In other words, the operator $p_{k}^{\dagger}$ acting on the state
$| \vec{n} \rangle$ with the configuration $\{ \vec{n} \}$ of bare phonons,
gives the eigenstate $| k ; \vec{n} )$ which contains one polaron of
momentum $k$ and the configuration $\{ \vec{n} \}$ of renormalized
phonons.
Defining ${\cal B}_{q} = e^{S} b_{q} e^{-S}$, we therefore obtain
${\cal B}_{q}^{\dagger} {\cal B}_{q} | k ; \vec{n} ) = n_{q} | k ; \vec{n} )$.
This is the straightforward generalization of the results obtained for the
two site system.

Using second order perturbation theory the polaron dispersion is 
given by
\begin{equation}
E(k) \; = \; \varepsilon_{k} \; + \; \frac{g^{2} \omega_{0}^{2}}{M} 
\sum_{q} \frac{1}{\varepsilon_{k} - \varepsilon_{k+q} - \omega_{0}}
\end{equation}
This is a well known result obtained by many approaches\cite{Mott,Mahan}.
This dispersion relation corresponds to a state with one polaron of
momentum $k$ and zero renormalized phonons. 
The next excited states correspond, for example, 
to states with one polaron of momentum $k$ and one renormalized phonon 
of momentum $q$.
The total momentum of these states is therefore $K=k+q$. The energy of such
a state is given by
\begin{equation}
E(k;q) \;= \; E(k) \;  + \omega_{0} 
\; + \; \frac{g^{2}\omega_{0}^{2}}{M} \left[ \frac{1}{\varepsilon_{k} 
- \varepsilon_{k+q} + \omega_{0}} + \frac{1}{\varepsilon_{k} 
- \varepsilon_{k-q} - \omega_{0}} \right]
\end{equation}
As concern the corresponding unitary transformation, it has been 
discussed by Fr\"ohlich\cite{Frohlich}
and is given by 
\begin{equation}
S \; = \; - \frac{g \omega_{0}}{\sqrt{M}} \sum_{q} \left\{ 
\frac{b_{q}}{\varepsilon_{k} + \omega_{0} - \varepsilon_{k+q}} +
\frac{b_{-q}^{\dagger}}{\varepsilon_{k} - \varepsilon_{k+q} - \omega_{0}}
\right\} \; c_{k+q}^{\dagger} c_{k} 
\end{equation}

We now discuss the excitation spectrum of the system in the weak coupling
limit and consider a one dimensional system for simplicity 
with the bare electron
dispersion $\varepsilon_{k} = -2t \cos(k)$. 

Our basic asumptions are a) that the dispersion of the polaron
$E(k)$ is slightly renormalized from the bare electron dispersion relation 
$\varepsilon_{k}$ and
b) the system is large enough so that the phonon frequencies are
slightly renormalized, so that $E(k;q) \simeq E(k) + \omega_{0}$. 
The ground state 
corresponds to the state with a single polaron of momentum $k=0$, 
with energy $E(0)$. The state of momentum $K$ with the lowest energy 
can be either the state with one polaron of momentum $k=K$ and no 
renormalized phonons or either the state with one polaron of 
momentum $k=0$ and one renormalized phonon of momentum $q=K$ such that 
$K=k+q$. The corresponding energies are respectively $E(K)$ and 
$E(0) + \omega_{0}$. 
Within our asumptions on $E(k)$, 
there exists a threshold $K^{*}$ such that $E(0) + \omega_{0} = E(K^{*})$.
For $K < K^{*}$ the lowest energy is $E(K)$ and the state contains only one
polaron of momentum $k=K$. For $K > K^{*}$ the lowest energy is
$E(0) + \omega_{0}$ and the state contains one polaron of zero 
momentum and one renormalized phonon of momentum $q=K$. 
As far as the others excitations are concerned, their construction is 
straightforward.
If we compare this spectrum with the results of Ref. \cite{Stephan,Fehske}, 
where this spectrum was computed, 
we find a nice agreement in the case of a weak coupling. 
The flat part in the spectrum  is just the energy $E(0) + \omega_{0}$. 
We further notice that for $K < K^{*}$, these numerical results show that the 
polaron dispersion $E(k)$ is slightly renormalized from the bare
electron dispersion $\varepsilon_{k}$.

\section{Conclusion}
In conclusion, we have shown  that for any coupling, one can define 
a renormalized electron of momentum $k$ and some renormalized bosons of
momentum $q$, such that the spectrum is made up of these renormalized 
excitations. The renormalized operators can be obtained via an unitary 
transformation. This was worked out  explicitly in the weak coupling limit
using perturbation theory. 
Recent finite size studies support our approach.

In the crossover regime, where level crossings arise in the spectrum of
the Hamiltonian , it is not possible to define the unitary 
transformation pertubatively  and it is not clear 
how to make some quantitative predictions.

\acknowledgements
The author is indebted to H. Castella and K. Lantsch for 
useful discussions.

\begin{figure}
\caption{The dispersion of the polaron for a one dimensional 
lattice with $\omega_{0}=0.8$ and $g^{2}\omega_{0}=0.1$. Energies are
in unit of $t$, solid line is the bare electron dispersion, dotted line 
is the slight polaron dispersion and dashed line is the lowest excitation 
as discussed in the text.}
\label{Fig1}
\end{figure}


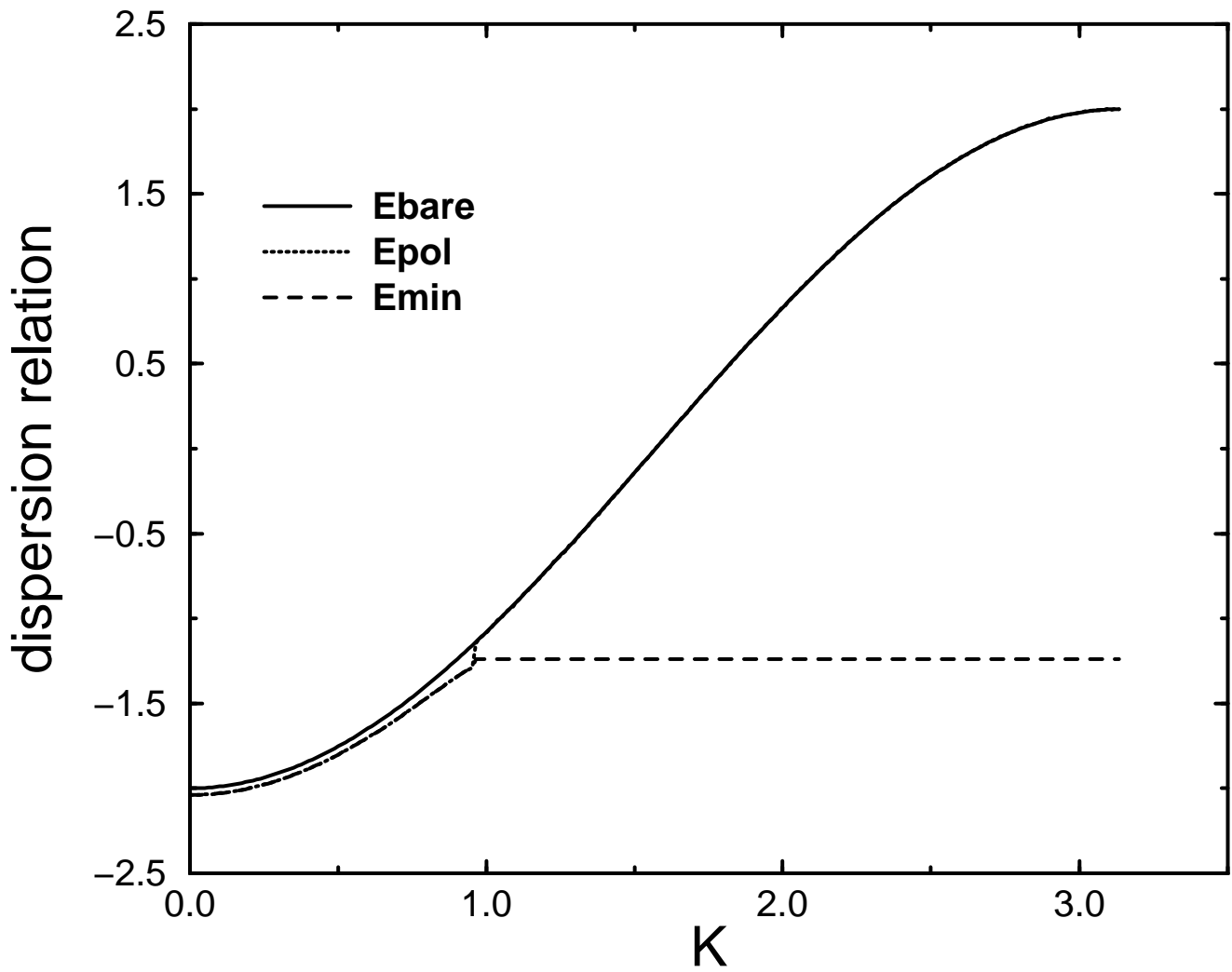
\end{document}